# *On the Theoretical Possibility of Quantum Visual Information Transfer to the Human Brain*

[1,2,3]**V. Salari**[*] , [1,2,3]**M. Rahnama**[†] and [4]**J. A. Tuszynski**[‡]

[1]*Department of Physics, Shahid Bahonar University of Kerman, P. O. Box 76175, Kerman, Iran*
[2]*Afzal Research Institute, Kerman, Iran*
[3]*Neuroscience Research Center, Kerman, Iran*
[4]*Department of Physics, University of Alberta, Edmonton, T6G 2J1, Canada*

***Abstract:***

The feasibility of wave function collapse in the human brain has been the subject of vigorous scientific debates since the advent of quantum theory. Scientists like Von Neumann, London, Bauer and Wigner (initially) believed that wave function collapse occurs in the brain or is caused by the mind of the observer. Experimentally, first Hall et al. performed an experiment to investigate wave function collapse caused by the mind of the observer. Their experiment did not detect any trace of wave function collapse as a result of human intentionality. A refined version of Hall et al.'s experiment was performed by Bierman in which a different result was obtained from what Hall et al. reported. On the basis of evoked potential diagrams, Bierman has concluded that brain can cause a collapse of external quantum states. It is a legitimate question to ask how human brain can receive subtle external visual quantum information intact when it must pass through very noisy and complex pathways from the eye to the brain? There are several approaches to investigate information processing in the brain, each of which presents a different set of conclusions. Penrose and Hameroff have hypothesized that there is quantum information processing inside the human brain whose material substrate involves microtubules (MTs) and consciousness is the result of a collective wavefunction collapse occurring in these structures. Conversely, Tegmark stated that owing to thermal decoherence there cannot be any quantum processing in neurons of the brain and processing in the brain must be classical for cognitive processes. However, Rosa and Faber presented an argument for a middle way which shows that none of the previous authors are completely right and despite the presence of decoherence, it is still possible to consider the brain to be a quantum system. Additionally, Thaheld, has concluded that quantum states of photons do collapse in the human eye and there is no possibility for collapse of visual quantum states in the brain and thus there is no possibility for the quantum state reduction in the brain. In this paper we conclude that if we accept the main essence of the above approaches taken together, each of them can provide a different part of a teleportation mechanism. Here, we propose a new model based on the premise that there exists a quantum teleportation mechanism between the eye and the brain. Specific assumptions used to build the model involve both classical and quantum mechanical elements. Our approach can combine the above seemingly contradictory conclusions in a compact and coherent model. This model revives this hypothesis that human brain can cause a collapse of quantum states, because in this model external quantum information can penetrate into the brain as an intact state.

## *1) Introduction*

Schrödinger's book "What is life?" has had an enormous influence on the development of molecular biology [1]. The great physicist's insight has inspired many researchers to investigate the molecular basis of living organisms [2],[3],[4]. Several researchers have noticed the sweeping consequences that would follow from the discovery that living organisms might process information quantum mechanically, either at the bio-molecular level, or the cellular/neuronal level [5],[6],[7],[8],[9]. Mainstream cognitive neuroscience has far largely ignored the role of quantum physical effects in the neuronal processes underlying cognition and consciousness. Clearly, many unsolved problems still remain, suggesting the need to consider new, possibly more radical approaches. These authors have proposed models in which the operation of consciousness is associated with some sort of explicit wave function collapse. There have been numerous suggestions that consciousness is a macroscopic quantum effect that may involve various physical phenomena associated with superconductivity, superfluidity, electromagnetic fields, Bose-Einstein condensation or some other physical mechanism [10,16,33]. Perhaps the most specific model developed thus far is that of Penrose and Hameroff and it asserts that quantum information processing takes place at the level of neuronal Microtubules (MTs). It has been argued that MTs can process information similarly to a cellular automaton, and hence Hameroff and Penrose suggest that neuronal MTs may operate as a quantum computer [17-19]. There are still open issues related to the persistence of quantum effects under physiological conditions, specifically the ambient temperature, but until conclusive experimental evidence is found for or against such effects, theoretical discussions will continue unabated.

---

[*] *vahidsalari@mail.uk.ac.ir , Tele Fax: 00983413222034*
[†] *Majid.rahnama@mail.uk.ac.ir*
[‡] *jtus@phys.ualberta.ca*

*2) Brain: Classical or Quantum Mechanical system?*

Classical physics is viewed by most scientists today as an approximation to the more accurate quantum theory, and therefore due to the nature of this classical approximation the causal effects of our conscious activity on the material substrate may appear to be eliminated.

One might well ask about the motivation for using quantum mechanics to explain different aspects of neuroscience. Here, we intend to discuss some of these motivations. Living systems are composed of molecules and atoms, and the most advanced theory for the explanation of the interaction between atoms and molecules is quantum theory. For example, making and breaking of chemical bonds, absorbance of frequency specific radiation (e.g. photosynthesis and vision), conversion of chemical energy into mechanical motion (e.g. ATP cleavage) and single electron transfer through biological polymers (e.g. DNA or proteins) are all quantum effects. Another reason is the *"binding" problem*. It means that we receive many sensory inputs at once: visual, auditory, olfactory, tactile and thermal [11]. The time intervals and locations of processing are different for each of them, but they interact with each other despite their relative distant locations and we perceive them as simultaneous events. This communication cannot be explained by conventional approaches adopted by neuroscience. Furthermore, Synaptic transmission and axonal transfer of nerve impulses are too slow to organize coordinated activity in large areas of the central nervous system. Numerous observations confirm this view [12]. The duration of a synaptic transmission is at least 0.5 ms, thus the transmission across thousands of synapses takes on the order of hundreds to thousands of milliseconds. The transmission speed of action potentials varies between 0.5 m/s and 120 m/s along an axon. More than 50% of the nerves fibers in the corpus callosum are without myelin, thus their speeds are reduced to approximately 0.5 m/s.[§] How can these low velocities (i.e. classical signals) explain the fast processing taking place in the nervous system? Moreover, the human body is made up of many organs, which themselves are made up of many billions of cells. How can such a system with billions of semi-autonomous components function effectively and coherently? [4] Explaining this is a major challenge since even relatively small-size human societies often undergo periods of turbulence and trouble due to conflict and poor organization. Now, some scientists think that quantum coherence is a major factor responsible for our bodies, and especially our brains, being so efficient and well coordinated. The above conclusions are sometimes invoked by the supporters of the quantum brain hypothesis.

---

[§] Updated from: http://highered.mcgraw-hill.com/sites/dl/free/0070960526/323541/mhriib_ch11.pdf, 2009-11-19.

Despite the potential power of quantum mechanics to answer the above questions, there are serious problems involved in considering it in the context of a living system. For instance, in order to have a very high degree of coherence between bio-molecules, Bose-Einstein condensation may be a viable effect, but we note that the ambient temperature in the human brain is too high for this phenomenon to occur. Now the question is "Can bio-molecules condense at this temperature or maintain coherence like lasers under these warm conditions?" Also, the sizes of bio-molecules and neurons are very large by physical standards to be regarded as typical quantum systems. Moreover, because of the noisy environment, according to decoherence theory, quantum states of these mesoscopic bio-molecules would collapse very rapidly. In addition, observation of quantum effects in living systems needs *very* accurate and sophisticated experimental instruments, and additionally, we note that it is very hard to extract information about the quantum phenomena occurring in the brain form complicated structures in this living system.

According to a number of recent papers published over the past decade [16,66,67] it appears that this conceptual challenge continues and the problem remains unsolved today. Here, we want to investigate this problem from a different point of view. Theoretically, we may consider the conscious observer of a quantum system and propose that the state of this system is reported via superposed photons. We address the question of whether the observer can receive the exact same state of this system quantum mechanically in his/her brain or this quantum state collapses before reaching the brain. Below, we investigate this problem in detail.

### *3) Evolution of Information from the Eye to the Brain*

We assume that a conscious observer directs his/her attention to a quantum system. For simplicity we consider this system to be a manifestation of the famous *Schrödinger's cat*. This system can exist in two states: Live $|L\rangle$ and Dead $|D\rangle$.

$$\psi_{sys} = \frac{1}{\sqrt{2}}(|L\rangle + |D\rangle) \qquad (2\text{-}1)$$

The state of this system is then reported via superposed photons. As is documented in the literature on the biophysics of vision, 4% of these photons are reflected from the cornea. 50% of the remaining photons are dissipated through ocular media absorption. The rest of the photons enter the 200-250 μm thick retina. There, they interact with the photoreceptors in the layer composed of rods and cons following an 80% loss due to retinal transmission [13,14]. In this case, we consider just a few remaining photons which are in a superposed quantum state. The expression "superposed quantum states" of the remaining photons is based on our first assumption that photons enter into the eye in superposed states, and also because the retina is largely transparent to photons so they can be received in the last layer of retina [15], Thus, we assume that the few remaining photons (which are not reflected, dissipated or absorbed) are still in a superposed state. The key question here is whether this quantum state of photons can be reported to the brain.

When this state interacts with the last layer of retina, it seems that this superposed photon undergoes a wavefunction collapse, because the photon's information signature will be converted into electrical signals after it leaves the retina. On the other hand, photons can be absorbed and then transformed into classical signals. Here, we use the symbols introduced by Tegmark [16] for the observer. The symbol $\left|\begin{smallmatrix}..\\-\end{smallmatrix}\right\rangle$

denotes the state for which the information on photons is not received by the brain and thus the observer is amphoteric. The symbol $\left|\overset{..}{\cup}\right\rangle$ stands for the state in which the information received in the brain reports that the cat is alive (and the observer is happy). Finally, the symbol $\left|\overset{..}{\cap}\right\rangle$ corresponds to the state in which the information received in the brain indicates that the cat is dead (and hence the observer is sad). It means that:

$$U\left|\overset{..}{-}D\right\rangle = \left|\overset{..}{\cap}D\right\rangle \qquad (2\text{-}2\text{-a})$$

$$U\left|\overset{..}{-}L\right\rangle = \left|\overset{..}{\cup}L\right\rangle \qquad (2\text{-}2\text{-b})$$

where $U = \exp\left[-\frac{i}{\hbar}\int H_{photon-brain}\,dt\right]$.

Now, we consider another state in which the brain interacts with itself. Penrose and Hameroff have proposed a model of consciousness involving quantum computation with objective reduction in *MTs* within the brain's neurons [17,18,19] (see Figure 1). MTs are cylindrical polymers comprised of the protein tubulin which organize numerous cellular activities including neuronal motor transport. According to Hameroff and Penrose, switching of tubulin conformational states is governed by quantum mechanical forces within the interior of each tubulin dimer, and an essential feature of the Orch OR model is that tubulin dimers may exist in quantum superpositions of two stable conformations. Therefore, these states could function as quantum bits, or "qubits" by interacting non-locally (through their entanglement) with other tubulin qubits so that MTs may act as quantum computers [17-19]. When sufficiently many entangled tubulins are superposed for a long enough time to reach Penrose's OR threshold given by E=h/T, where E is the gravitational self-energy of the system, h is Plank's constant and T is the decoherence time, an objective reduction (OR) "conscious event" occurs as stated in the Orch-OR model,

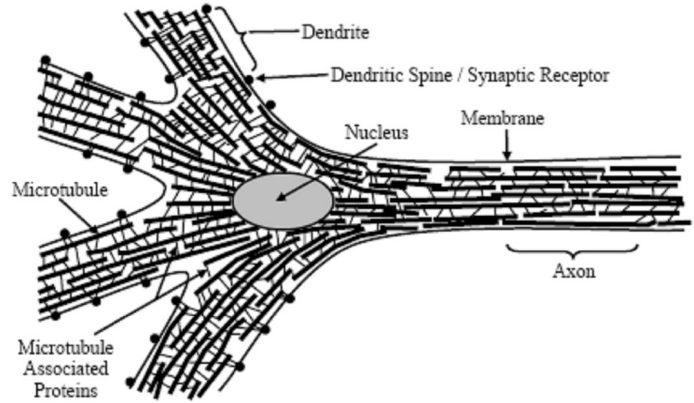

Figure 1- Representation of MTs in a brain neuron. The Orch OR model suggests that the main information processing is implemented in these structures.

If the previous evolution is described by Penrose's self-collapse in the brain (i.e. Orch-OR), MTs in the neurons of the brain collectively evolve and then collapse (i.e. undergo a conscious event) to one of the happy or sad states. It mathematically means that:

$$U \left| \begin{matrix} .. \\ - \end{matrix} \right\rangle = \frac{1}{\sqrt{2}} ( \left| \begin{matrix} .. \\ \cup \end{matrix} \right\rangle + \left| \begin{matrix} .. \\ \cap \end{matrix} \right\rangle ) \qquad (2\text{-}3)$$

where $U = \exp\left[-\frac{i}{\hbar} \int H_{brain} dt \right]..$

According to the Orch OR model, consciousness is due to an objective reduction or self collapse in the brain; however, we know that consciousness can be attributed mainly to the interaction of external information with bio-structures inside the brain. Consequently, if we accept both of these conclusions simultaneously, we have to say that the outcome of consciousness should be identical for both of the above conclusions. This is impossible unless we propose that retina and brain are strongly correlated or entangled with each other. If we compare the state in Eq. (2-3) and the state which has the information about the photon in Eq. (2-2), we can say that there is a great amount of correlation between the retina and the visual cortex, because the results registered by them should be identical. An additional argument for this correlation is that in accordance with the Einstein-Podolsky-Rosen (EPR) pair, when two entities originate from a common source they can be entangled with each other [20,21]. Retina has a similar layered structure as the top layers of the gray matter in the cerebral cortex of the brain [22]. In fact, retina is an extension of the central nervous system (the brain and spinal cord) that forms during embryonic development [22]. One reason why scientists are interested in retinal processing is that retina is an accessible part of the brain that can be easily stimulated with light [22]. To better explain 'nonlocal' correlations between the visual cortex and retina, we discuss a simple example in connection with another of the five senses. The sense of touch also involves nonlocal correlations between the brain and receptor cells. Suppose we prick the top of our finger with a needle which produces a single action potential to convey/report the sense of pain to the brain. There is no sensation of pain in the finger before receiving the action potential in the brain. As soon as the action potential is received in the brain we feel the sensation of pain in our finger[**]. The important question is: where is the site of consciousness here (i.e. the sensation of pain)? Is it in the brain or in the finger? The signal is in the brain (i.e. not in our finger anymore) BUT the sensation of pain is in the finger. There should be a mechanism involving an instantaneous interaction between the brain and receptor cells, because as soon as the signal is received in the brain a process takes place triggering the sensation of pain in the finger. It is impossible to explain it using classical interactions. Even if we assume that after receiving an action potential in the brain a feedback signal will be produced and transmitted to the site of injury, then it takes approximately a second to propagate it, which is not consistent with what really happens [12]. If we assume that this fast response is due to electromagnetic interaction generated in the brain, this possibility is not easily acceptable because the brain is an organ formed by large numbers of cell membranes with high impedance and a small amount of extracellular fluid in between. Electrical potentials produced by neurons and by glia must pass through high-impedance cell membranes and hence cannot travel too far without being absorbed or dissipated. For instance, the electrical potentials recorded in an electroencephalogram originate in the most superficial layers of the cerebral cortex [12]. Potentials from deeper structures can be recorded only after numerous repetitions of the sweeps and their averaging, as seen in the recording of auditory brainstem potentials. Also, when unit activity is recorded extra-cellularly, it is difficult to extract a signal from the noise at a distance larger than 100 microns. There must be a mechanism that acts faster than synaptic transmission and that is different from electromagnetic interactions [12]. Thus, based on the above arguments, we can venture to state that retina and the visual cortex are entangled with each other. Admittedly, it is hard to say which bio-molecular structures in the body can be responsible for entanglement between the brain and some other part of the body. We believe MTs are good candidates for providing this type of entanglement because they exist in all cells of the body and can be viewed similarly to the wiring of a building. MTs are particularly numerous in the brain where they form highly ordered bundles [19]. MTs can be electrically polarized-depolarized on the order of ns and even ps due to the motions of highly charged protrusions called C-termini [23]. Thus they may be able to create coherent states for unlimited times, so the entangled coherent states can be recreated countless times between the visual cortex and retina.

Centrioles and cilia, which are complex microtubular structures, are involved in photoreceptor functions in single cell organisms and primitive visual systems. Cilia are also found in all retinal rod and cone cells. The dimensions of centrioles and cilia are comparable to the wavelengths of visible and infrared light [24]. In a series of studies spanning a period of some 25 years of research G Albrecht-Buehler (AB) demonstrated that living cells possess a spatial orientation mechanism located in the centriole [25, 26, 27]. This is based on an intricate arrangement of MT filaments in two sets of nine triplets each of which are perpendicular to each other. This arrangement provides the cell with a primitive "eye" that allows it to locate the position of other cells within a two to three degree accuracy in the azimuthal plane and with respect to the axis perpendicular to it. He further showed that electromagnetic signals are the triggers for the cells' repositioning. It is still largely a mystery how the reception of electromagnetic radiation is accomplished by the centriole. Moreover, the cytoskeleton is found mostly among the retina and the visual cortex in the cells of the optic nerve as is found in all nerve cells. Cytoskeletal structures of the centrioles can be expected to vibrate like a harmonic oscillator in its ground

---

[**] Updated from http://www.mydr.com.au/pain/pain-and-how-you-sense-it 28 November 2010

state. Vibrational dynamics of MT's has been the subject of a recent paper where typical frequency ranges have been discussed [28].

When photons interact with a centriole, their electric field can displace the potential of the harmonic oscillator and then releases it, thus generating coherent states [29]. We denote here $|z\rangle$ as a coherent state: $|z\rangle = \exp(-\frac{|z|^2}{2})\sum_{n=0}^{\infty}\frac{z^n}{\sqrt{n!}}|n\rangle$, where z is the eigenvalue of the annihilation operator.

Hameroff [30] and Penrose [17] have suggested that MTs inside cells support long-range quantum coherence, enabling quantum information processing to take place at the sub-cellular level. They use this hypothesis to develop their theory of consciousness. Cells interconnected by gap junctions form networks which fire synchronously, behaving like one giant neuron [31], and possibly accounting for synchronized neural activity such as coherent 40 Hz waves [32]. Marshall [33] has suggested that coherent quantum states known as Bose-Einstein condensates occur among neuronal proteins [34],[35],[36]. Other issues, such as preconscious-to-conscious transitions were identified and discussed by Stapp [37] with the collapse of a quantum wave function in pre-synaptic axon terminals. The other reason for coherence of these structures is that light is an electromagnetic wave and thus represents a vibrational degree of freedom. According to Fröhlich's theory [38,39,40], it can excite within these cytoskeletal structures (i.e. nonlinear structures composed of dynamic electric dipoles) a single mode of frequency giving rise to long-range coherence.

Centrioles are two mutually perpendicular cylinders each of which is composed of nine MT triplets surrounding a central axoneme (which, according to Hameroff may be of significance in the molecular origin of cancer [41]). We assume when a coherent state $|z\rangle$ is generated in one centriole, in the other it will generate the state $|-z\rangle$. Now, we can say that after the interaction of photons with centrioles, they cause centrioles to vibrate and generate "entangled coherent states" [41] in these structures in the retina, i.e.:

$$|\phi\rangle_{first\,state_{12}} = (A|z\rangle_1|z\rangle_2 \pm B|-z\rangle_1|-z\rangle_2) \quad\quad (3\text{-}2)$$

where A and B are coefficients and $|\phi\rangle_{first\,state_{12}}$ is an entangled coherent state in centrioles with two modes 1 and 2. Here, our intention is not to convince the reader to accept that this type of entanglement crucially exists in centrioles, but we hypothetically accept the assumption of Hameroff [41] and formulate it in terms of mathematics for our next set of calculations.

The QED-cavity model of MTs [42] asserts that coherent modes of electromagnetic radiation can be sustained in the interior of MTs. These modes are provided by the interaction of the electric dipole moments of the ordered-water molecules in the interior of MTs with quantized electromagnetic radiation [43,44]. Jibu, et al. [45] have proposed that the quantum dynamical system of water molecules and the quantized electromagnetic field confined inside the hollow MT core can manifest a specific collective dynamical effect called superradiance [46] by which the MT can transform any incoherent, thermal and disordered molecular, atomic or electromagnetic energy into coherent photons inside the MT. Furthermore, they have also shown that such coherent photons created by superradiance penetrate perfectly along the internal hollow core of the MT as if the optical medium inside it were made "transparent" by the propagating photons themselves. This is referred to as the quantum phenomenon of self-induced transparency [47]. Superradiance and self-induced transparency in cytoskeletal MTs can lead to "optical" neuronal holography [48]. Thus Jibu, et al. [45], suggest that MTs can behave as optical waveguides which result in coherent photons. They estimate that this quantum coherence is capable of superposition of states among MTs spatially distributed over hundreds of microns. These in turn are in superposition with other MTs hundreds of microns away in other directions, and so on. With the above conclusions $|\phi\rangle_{first\,state\,12}$ can produce those photons which produced themselves, thus if the state $|\phi\rangle_{first\,state\,12}$ can be restored in the brain, it will reproduce the photons which were absorbed in the retina.

Additional arguments in favor of the feasibility of photon production in the brain can be found in the conclusions of the papers by Bokkon [49,50], who also asserts that there exists a neural activity-dependent ultra-weak photon (biophoton) emission in the brain. Thus there is the possibility to restore the initial state of the photon in the brain after absorption in the eye. This process can be implemented through the quantum teleportation mechanism between the retina and the visual cortex as will be discussed in the following sections.

Recent advances in femtosecond laser-based two-dimensional spectroscopy and coherent control have made it possible to directly determine the relevant timescales of quantum coherence in biological systems and even manipulate such effects. The picture that is emerging is that there are primary events in biological processes that occur on timescales commensurate with quantum coherence effects [51]. In a recent landmark paper, Sension [52] presented convincing arguments showing that plants and bacteria harvest light for photosynthesis so efficiently because of the coherent application of quantum principles.

### 4) *MTs, Coherence and Decoherence Issues*

As we discussed before, the reason for coherence of biomolecules in neurons, especially in MTs, is that light is an electromagnetic wave and thus represents a vibrational degree of freedom. According to Fröhlich's theory [38-40] it can excite within these microtubular structures (i.e. nonlinear structures composed of electric dipoles) a single frequency mode giving rise to long-range coherence. The Wu-Austin Hamiltonian [53,54,55] was proposed to describe the interaction of quantized electromagnetic field with a dipolar system to give a coherent Fröhlich's state. This Fröhlich condensation is used as a quantum coherent state for a biological system. There are different approaches possible to the coherent state generation in biological systems based on Fröhlich's coherent states as described in the works of Mequita et al. [56,57,58]. Bolterauer and Ludwig [59] investigated the thermodynamics of Wu and Austin system quantum mechanically and have shown that even without pumping their Hamiltonian can give rise to Bose condensation. The Wu–Austin Hamiltonian has the unphysical property of having no finite ground state [60]. Turcu [61]have obtained a master equation for Fröhlich rate equations. The main aim of his work was to show that there is a rich family of Hamiltonians, modeling differently the pump and the thermal bath, from which the same Fröhlich like rate equations can be obtained. We believe that the system of neuronal MTs is a good candidate for being described by one of these Hamiltonians. MTs are composed of tubulins which can be considered as biological electric dipoles. Pokorny provided a detailed analysis of the coherent states in MTs. He experimentally observed resonance effects in MTs in the range of MHz [62].The following is the Wu-Austin Hamiltonian for a biological system composed of electric dipoles with *N* modes connected to harmonic baths in exposure to a quantized electromagnetic source for $N_B$ relaxation-bath modes *k* of frequency $\Omega_k$ and $N_I$ input electromagnetic modes *l* of frequency $\Omega_l^{'}$

$$H_{Wu-Austin} = \sum_{i=1}^{N} \hbar\omega_i a_i^+ a_i + \sum_{k=1}^{N_B} \hbar\Omega_k b_k^+ b_k + \sum_{l=1}^{N=I} \hbar\Omega_l^{'} c_l^+ c_l + \sum_{i=1}^{N}\sum_{l=1}^{N_I}(\gamma a_i c_l^+ + \gamma^* a_i^+ c_l)$$
$$+ \sum_{i=1}^{N}\sum_{j=1}^{N_B}\sum_{k=1}^{N_B}(\alpha a_i b_j^+ b_k + \alpha^* a_i^+ b_j b_k^+) + \sum_{i=1}^{N}\sum_{j=1}^{N}\sum_{k=1}^{N_B}(\beta a_i a_j^+ b_k + \beta^* a_i^+ a_j b_k^+)$$

(4-1)

where $a_i$, $b_k$ and $c_l$ are creation operators for the system, heat bath, and quantized electromagnetic field, respectively. Pokorny and Wu have generalized the Wu-Austin Hamiltonian by considering new nonlinear terms added to the Wu-Austin Hamiltonian [63]. In the Wu-Austin Hamiltonian the vibration of the system is linear and its coupling with an external source and the heat bath is nonlinear. However, the vibration spectrum of the system may have nonlinear properties without interactions with the heat bath. It was shown that strong electric fields in proteins, membranes, and cytoskeleton polymers can be responsible for such nonlinearity [63]. The Hamiltonian proposed is

$$H = H_0 + H_1 + H_2 \qquad (4-2)$$

Where

$$H_0 = \sum_{i=1}^{N} \hbar\omega_i a_i^+ a_i + \sum_{k=1}^{N_B} \hbar\Omega_k b_k^+ b_k + \sum_{l=1}^{N=I} \hbar\Omega_l^{'} c_l^+ c_l \qquad (4-3)$$

which contains the first three terms of the Wu-Austin Hamiltonian in equation (4-1). The next two terms are

$$H_1 = \sum_i\sum_k\left(\varphi a_i b_k^+ + \varphi^* a_i^+ b_k\right) + \sum_i\sum_l\left(\zeta c_l a_i^+ + \zeta^* c_l^+ a_i\right) \qquad (4-4)$$

$$H_2 = \sum_{j,l,p}\sum_{m,n}\left\{\begin{array}{l}\left[\sigma a_j^+ (a_l)^m (a_p)^n + \sigma^* a_j (a_l^+)^m (a_p^+)^n\right]\\ +\left[\rho a_j^+ (a_l)^m (a_p)^n + \sigma^* a_j (a_l^+)^m (a_p^+)^n\right]\\ +\left[\xi a_j^+ (a_l)^m (a_p^+)^n + \xi^* a_j (a_l^+)^m (a_p)^n\right]\end{array}\right\} \qquad (4-5)$$

The perturbation $H_1$ in equation (4-3) contains two terms in which $\varphi$ and $\zeta$ are coupling constants for linear coupling to the heat bath and the source of energy, respectively. In equation (4-5), $\sigma$, $\rho$, and $\xi$ are the coupling constants and symbols of the type $(a_i)^m$ stand for $a_i a_i a_i ... a_i$ applied m-times.

Previously, one of the concerns regarding coherent states in the brain involved the fact that the Bose-Einstein condensation typically occurs only at low enough temperatures, much higher than body temperature. Recently, Reimers et al. [64] have argued that a very fragile Fröhlich coherent state may only happen at very high temperatures and thus there is no possibility for the existence of Fröhlich coherent states in biological systems, so every quantum model based on Fröhlich coherent state should be ruled out. it has also been shown that there are serious problems in their calculations and consequently their conclusions appear not to be credible [65].

The key question about the potential for quantum information processing in MTs is: "how is it possible for MTs to process information quantum mechanically while the environment surrounding them is relatively hot, wet and noisy?"

According to the decoherence theory, macroscopic objects obey quantum mechanics. The interaction with the environment in this theory causes decoherence, which destroys quantum effects of macroscopic objects [67]. Tegmark [16] has calculated decoherence times for MTs based on the collisions of ions with MTs leading to the decoherence times on the order of:

$$\tau = \frac{D^2 \sqrt{mkT}}{Ngq^2} \approx 10^{-13} s \qquad (4\text{-}6)$$

where $D$ is tubulin diameter, $m$ is the mass of the ion, $k$ is Boltzmann's constant, $T$ is temperature, $N$ is the number of elementary charges in the MT interacting system, $g = \frac{1}{4\pi\varepsilon_0}$ is the Coulomb constant and $q$ is the charge of an electron. According to Hagan et al.[66], Tegmark's interpretation is not aimed at an existing model in the literature but rather at a hybrid that replaces the superposed protein conformations of the Orch. OR theory with a soliton in a superposition state along the MT. Another main objection to this estimate is that Tegmark's formulation yields decoherence times that increase with temperature contrary to well-established physical intuitions and the observed behavior of quantum coherent states. In view of these (and other) problems in Tegmark's estimates, Hagan et al. [66] assert that the values of quantities in the Tegmark's relation are not correct and thus the decoherence time should be approximately $10^{10}$ times greater. Rosa and Faber [67] have also revised Tegmark's formulation and obtained their relation as:

$$\tau = \frac{1}{gq_1q_2} \frac{\hbar^3}{x_1 MkT} \qquad (4\text{-}7)$$

Where $q_1$ and $q_2$ are electrical charges of tubulin and environmental ions, respectively, and $x_1$ is the x component of the tubulin distance to the origin of their proposed coordinate set. This formulation is very different from what Tegmark has obtained. The formulation of Rosa and Faber shows that the decoherence time becomes too high when temperature is very low, and it is compatible with what exists in the quantum computation literature. It thus appears that this problem is not resolved yet and there is no general relation between the decoherence time and temperature. For example, lasers maintain their coherence at high temperatures due to external pumping. Moreover, quantum spin transfer between quantum dots connected by benzene rings (the same structures found in aromatic hydrophobic amino acids) is more efficient at warm temperature than at absolute zero [68]. Tegmark defends his formulation [69] and believes that the point Hagan et al. [66] overlooked is that as soon the absolute temperature is lowered by about 10%, below 0 Celsius, the brain freezes and the decoherence time grows dramatically. The slight decrease in decoherence time for tiny temperature reductions simply reflects the fact that the scattering cross-section grows as the temperature is lowered, just as slow neutrons have larger cross section than fast ones in a nuclear reactor.

According to the Orch-OR model, MTs in neurons of the brain process information quantum mechanically and they avoid decoherence via several mechanisms over sufficiently long times for quantum processing to occur. According to the Orch OR model, MTs like lasers maintain quantum coherence against thermal noise. Water within cells is itself not truly liquid, but has been shown to be, to a large extent, ordered [70]. Most of the ordered water in the cell in fact surrounds MTs [71]. MTs and other cytoskeletal components are embedded in cytoplasm which exists in alternating phases of (1) "sol" (solution, liquid); and (2) "gel" (gelatinous, "solid"). Among the most primitive of biological activities, "sol-gel transformations" within neurons and other living cells are caused by assembly and disassembly of cytoskeletal actin (e.g. regulated by calcium ions through the protein calmodulin, in turn regulated by MTs). Sol-gel transformations are essential in basic cellular activities such as ("amoeboid") movement, growth and synaptic formation and neurotransmitter vesicle release [72,73]. Transitions can occur rapidly (e.g. 40 sol-gel cycles per second), and some actin gels can be quite solid, and withstand deformation without transmitted response [74]. Cyclical encasement of MTs by actin gels may thus be an ideal quantum isolation mechanism. In the gel phase of cytoplasm, the water ordering surfaces of a MT are within a few nanometers of actin surfaces which also order water. Thus bundles of MTs encased in actin gel may be effectively isolated extending over the radius of the bundle, on the order of hundreds of nanometers. There are many mechanisms which can protect these structures against decohering factors. In general, quantum states of tubulin/MTs may be protected from environmental decoherence by biological mechanisms which include phases of actin gelatin, plasma-like Debye layering, coherent pumping and topological quantum error correction [66]. MTs may possibly utilize nonspecific thermal energy for "laser-like" coherent pumping, for example in the GHz range by a mechanism of "pumped phonons" suggested by Fröhlich [38-40].

### 5) Can quantum states of environmental photons be restored in the brain?

Recently research in quantum state transfer, especially in quantum teleportation, has emerged as one of the major research areas of theoretical and experimental quantum mechanics [75]. There is a simple scheme of quantum teleportation in Fig. 2. Assume that Alice wants to send Bob an unknown quantum state, but, when she receives this state, she does not know anything about that, unless she affects it and collapses it to a classical state, or in other words she destroys that quantum state. She can just send classical signals to Bob through a classical channel, but if there is a shared entangled channel between Alice and Bob, Bob can reconstruct the initial quantum state with the help of a classical signal which is sent by Alice and a quantum channel between them. This operation is implemented by the use of special unitary operators. Now we explain the teleportation mechanism between Alice and Bob according to [75]. Assume that there is an entangled pair between Alice and Bob, one part is on behalf of Alice and the other for Bob. This pair is shown as $|\beta_{xy}\rangle = \dfrac{|0,y\rangle + (-1)^x |1,\bar{y}\rangle}{\sqrt{2}}$ in which $x$ and $y$ take the numbers 0 and 1, so there are four possible values for $|\beta_{xy}\rangle$ as $|\beta_{00}\rangle, |\beta_{01}\rangle, |\beta_{10}\rangle$, and $|\beta_{11}\rangle$. These states are known as Bell states, or sometimes the EPR states or EPR pairs [75].

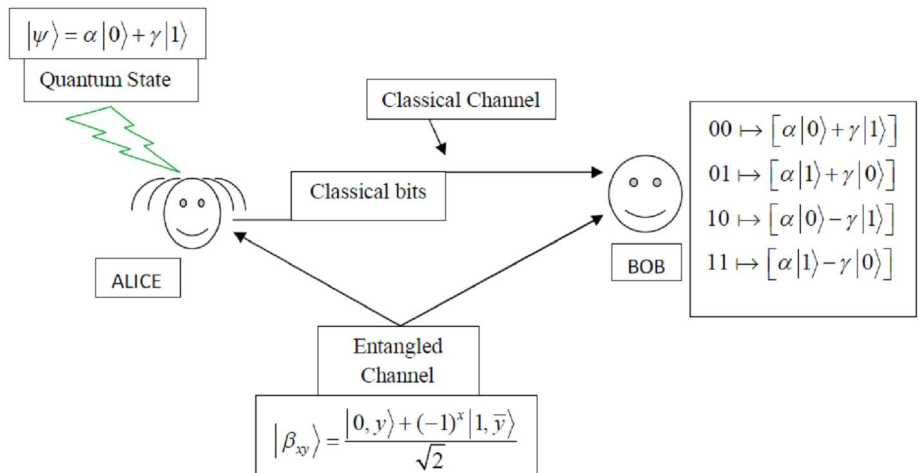

Figure 2- A typical scheme of Teleportation between Alice and Bob. In this scheme the quantum state $|\psi\rangle = \alpha|0\rangle + \gamma|1\rangle$ is teleported from Alice to Bob.

The state to be teleported is $|\psi\rangle = \alpha|0\rangle + \gamma|1\rangle$. The amplitudes $\alpha$ and $\gamma$ are unknown. Alice interacts with the state $|\psi\rangle$ with her half of the EPR pair which is $|\psi\rangle|\beta_{xy}\rangle$, and then measures the two qubits in her possession, obtaining one of four possible classical results 00, 01, 10, and 11. She sends this information to Bob. Depending on Alice's classical message, Bob performs one of four operations on his half of the EPR pair. For example in the case where the measurement yields 00, Bob doesn't need to do anything. If the measurement is 01 then Bob can fix up his state by applying the *X* operator, $X \equiv 1|0\rangle\langle 1| + 1|1\rangle\langle 0|$. If the measurement is 10 then Bob can fix up his state by applying the *Z* operator, $Z \equiv 1|0\rangle\langle 0| - 1|1\rangle\langle 1|$. If the measurement is 11 then Bob can fix up his state by applying *ZX* which means first an *X* and then a *Z* operator. Amazingly, by doing each operation he can recover the original state $|\psi\rangle$ [75].

Here, we would like to simulate visual quantum information transfer with the help of a quantum teleportation mechanism. In this paper we use coherent states $|z\rangle$ instead of the kets $|0\rangle$ and $|1\rangle$. The operators X, Z and ZX will be replaced with 'phase shift' operators ($e^{-i\pi a_i^\dagger a_i}$ and $e^{-i\pi(a_i^\dagger a_i + a_j^\dagger a_j)}$ *(i,j=1,...,6)* where $a^\dagger$ and *a* are creation and annihilation operators, respectively) which are created naturally by a phase difference between coherent states in neurons (see section 6). We know that when photons penetrate the retina, they change to action potentials or electrical signals and these classical signals are sent to the brain for interpretation. It means that *retina* (Alice) wants to send the *brain* (Bob) a *photon state* (unknown quantum state which she has received), but *retina* (Alice) absorbs it (collapses the quantum state) and changes it to an *action potential* (classical state) and sends it through *membranes of the axons of brain neurons* (classical channel). *Brain* (Bob) can reconstruct the initial *state of the photon* (unknown quantum state) to process it resulting in the emergence of consciousness.

In summary, our argument for the quantum teleportation mechanism which uses all the major arguments offered in this connection before is as follows:
1- *According to Orch OR*: There is quantum information processing taking place in the neurons of the brain (there is a quantum channel between retina and the brain)
2- *According to Tegmark*: Displacement of ions through membranes of brain neurons is a classical phenomenon (action potentials are classical signals and membranes of neurons are classical channels).
3- *According to Rosa and Faber:* despite decoherence, brain can be considered to be a quantum processor.
4- *According to Thaheld*: Superposed photons do collapse in the retina (the quantum state is collapsed by the sender [Alice]).

Note that the above four steps are all correct, but not indepenedtly and only as parts of the quantum teleportation process which we suggest in this paper (see Table 1).

*Table 1 Simulation of visual information from the eye to the brain via quantum Teleportation mechanism.*

| "Human Brain" | Quantum Teleportation Mechanism |
|---|---|
| Retina | Alice |
| Membrane of axons in the neurons | Classical channel |
| Cytoskeletal structures | Entangled channel (quantum channel) |
| Visual cortex | Bob |
| Action potentials | Classical signals |

It is worth noting that a model of teleportation in MTs was proposed earlier by Mavromatos et al. [42] but it does not explain the role of classical signals and action potentials in neuronal signaling.

Now we would like to investigate in more detail this teleportation mechanism via entangled coherent states through visual pathways. We will show how photon states can be constructed in the visual cortex.

## 6) The Plausibility Arguments for Teleportation of Entangled Coherent States through Visual Pathways

As we explained before, Superradiance and self-induced transparency in addition to Fröhlich's dipolar oscillations can cause the coupling of MT dynamics over long distances and create a superposition coherent state. While in a superposition state, tubulin dimers may mutually communicate in the same manner, and in MTs in neighboring neurons, and through macroscopic regions of the brain via tunneling through gap junctions and possibly tunneling nanotubes [30]. As mentioned before, retina and the visual cortex can be entangled with each other. Thus, there can be a quantum channel between retina and the visual cortex which is composed of microtubular structures. MTs interact with membrane structures mechanically by linking proteins, chemically by ions and second messenger signals, and electrically by voltage fields. Transduction of light into electrical signals takes place in the photoreceptors [76].

Axons leaving the temporal half of the retina traverse the optic nerve to the optic chiasm, where they join the optic tract and project to ipsilateral structures. Axons leaving the nasal half of the retina cross the midline at the chiasm and terminate in contralateral structures [76]. This arrangement means that all the axons in the optic tract carry information about the contralateral visual field. Axons of the optic tract terminate in three areas of the central nervous system, the lateral geniculate nucleus (i.e. LGN), the superior colliculus and the pretectal area. The trajectory through the LGN is the largest most direct and clinically most important pathway by which visual information reaches the cerebral cortex [76]. About 80% of the optic tract axons synapse in the LGN. The LGN is a laminated structure, having 6 layers. Contralateral fibers and ipsilateral fibers couple in the LGN (see Fig. 3). The ipsilateral fibers of the optic nerve terminate in laminae 2,3 and 5 of LGN, while the contralateral fibers terminate in laminae 1, 4 and 6 of LGN [76]. There are about $10^6$ neurons in each LGN, all of which project to the ipsilateral occipital cortex (area 17) as the optic radiations. The portion of the cerebral cortex that receives LGN axons is called the striate cortex and is usually labeled V1 to designate it as the primary visual cortical area (see Figure 2). Virtually all information in the visual system is recognized as being processed by V1 first, and then passed on to higher order systems [76, 77].

Now, we investigate the information transfer through visual pathways. As we discussed before the Orch OR model asserts that the main information processing in the neurons of the brain is performed in the MTs and the nature of the processing is mainly quantum mechanical. The processing unit in this model is *tubulin* which can be in a superposed state. Tubulins act like qubits in quantum computers. *Tegmark* has vigorously argued against quantum processing in the human brain having calculated decoherence times for every superposition state possible in the neurons of the brain [16]. In his opinion, superposition states include ions such as $Na^+$ which are "in" and "out" of the membrane of an axon. On the other hand, $Na^+$ ions are in the superposition of "in" and "out" with a separation distance comparable to the membrane thickness. He has considered three factors which can destroy this superposition state in neurons. *Collisions with the neighboring ions, collisions with the water molecules* and *interactions with distant ions* are the factors which Tegmark investigated for decoherence. He estimated the corresponding decoherence times to be in the range between $10^{-19}$ s and $10^{-20}$ s. It is clear that these decoherence times are extremely small on the time scale of the brain processes such as seeing, thinking, speaking and the other cognitive processes. Typically, dynamical timescales for neuron firing and cognitive processes are in the range of $10^{-4}$ to 1 second, whereas decoherence timescales are many orders of magnitude shorter. Thus, action potentials should be regarded as classical signals and the displacement of ions through the membrane of axons should be investigated classically. It is worth noting that Tegmark has also calculated decoherence times for MTs, but these calculations were made under inappropriate assumptions about these structures (for more details see [66]) and hence, while we can accept his calculations about action potentials, the calculations for MTs appear not to be relevant to the problem discussed here. Rosa and Faber [67] have corrected the Tagmark decoherence time formula for MTs and have asserted that if we replace the gravitational collapse of Orch OR model with decoherence the quantum approach to brain problem remains strong. Thaheld [13-14,78] asserts that the wave function of any superposed photon state or states is always objectively changed within the complex architecture of the eye, and any incident photons have to run a very daunting gauntlet before they are even converted or transduced to retinal ganglion cell spike trains (to learn more about Thaheld arguments, the reader is referred to reference [79]). According to Thaheld, the quantum state of photons does collapse in the retina and it does not reach the brain. Is Thaheld really right? Is not there any mechanism to rebuild the quantum state of photons in the brain? Here, we accept that the states of photons collapse in the retina but we believe that they can be restored in the visual cortex via the teleportation mechanism described above.

Now, the question is: "how can it be possible to restore the exact state of photons in the brain while their state is collapsed in the retina?" The other question which one may ask is: "if this state is reported through action potentials how is this information reported to the brain and how can it interpret action potentials to obtain the exact state of the photons?" Our solution to the above problems involves the teleportation of entangled coherent states through visual pathways. The state of the photon is teleported from the eye to the brain. On the other hand, the state of the photon is transferred via some "cut-and-paste" mechanism from the eye to the brain. But how is it possible?

We explained before that retina and the visual cortex are entangled. Also we explained how the entangled coherent state is generated in the retina. Now, we wish to formulate the process of information transfer from the retina to V1. The state (3-2) with two modes 1 and 2 should be teleported to V1. After the interaction of light with retina, modes 3, 4 and 5, 6 are generated through microtubular structures between retina and V1, and thus they can produce entangled coherent channels between retina and V1. It means that the channels are:

$$|\psi\rangle_{\text{Retina}-V1_{35}} = \delta_3 |z\rangle_3 |z\rangle_5 \pm \delta_4 |-z\rangle_3 |-z\rangle_5 \qquad (6\text{-}1)$$

$$|\phi\rangle_{\text{Retina}-V1_{46}} = \delta_5 |z\rangle_4 |z\rangle_6 \pm \delta_6 |-z\rangle_4 |-z\rangle_6 \qquad (6\text{-}2)$$

where $\delta_i$ (i=3,4,5,6) are coefficients. We let minus signs for $\delta_4$ and $\delta_6$ for simplicity in subsequent calculations. Each mode is reported via a special fiber through visual pathways. All of the neurons which are collected in the LGN are divided into two major pathways: ipsilateral fibers and contralateral fibers. Information transfer in the contralateral fibers takes longer than information transfer in ipsilateral fibers because contralateral fibers have crossing relative to ipsilateral fibers and then they have longer lengths than ipsilateral fibers. On the other hand, contralateral fibers have a retarded phase relative to ipsilateral fibers. Now we will attempt to answer the following questions. What is this phase difference? What is the role of this crossing? And how does crossing restore the initial state in the retina?

### 7) *The Role of Phase Shift to Restore Information in LGN*

When the information is collapsed in the retina, action potentials are produced. The shape of action potentials is the same for each neuron, but the main problem is which neurons are fired, or in other words which neurons carry action potentials and information. Consider two fibers selected from ipsilateral fibers and two fibers selected from contralateral fibers. The two ipsilateral fibers are denoted Latin numerals i and ii, and the two contralateral fibers are denoted by iii and iv while the two fibers from the LGN to V1 are denoted by v and vi which are selected from the group of magnocellular and parvocellular fibers. Now, we start from the retina. The state of centrioles and channels is:

$$|\psi'\rangle_{LGN} = |\phi\rangle_{first\,state_{12}} \otimes |\psi\rangle_{Retina-V1_{35}} \otimes |\phi\rangle_{Retina-V1_{46}}$$

$$\begin{aligned}
&= A\delta_3\delta_5 |z\rangle_1|z\rangle_2|z\rangle_3|z\rangle_4|z\rangle_5|z\rangle_6 \pm A\delta_4\delta_5 |z\rangle_1|z\rangle_2|-z\rangle_3|-z\rangle_4|z\rangle_5|z\rangle_6 \\
&\pm A\delta_3\delta_6 |z\rangle_1|z\rangle_2|z\rangle_3|z\rangle_4|-z\rangle_5|-z\rangle_6 + A\delta_4\delta_6 |z\rangle_1|z\rangle_2|-z\rangle_3|-z\rangle_4|-z\rangle_5|-z\rangle_6 \\
&\pm B\delta_3\delta_5 |-z\rangle_1|-z\rangle_2|z\rangle_3|z\rangle_4|z\rangle_5|z\rangle_6 + B\delta_4\delta_5 |-z\rangle_1|-z\rangle_2|-z\rangle_3|-z\rangle_4|z\rangle_5|z\rangle_6 \\
&+ B\delta_3\delta_6 |-z\rangle_1|-z\rangle_2|z\rangle_3|z\rangle_4|-z\rangle_5|-z\rangle_6 \pm B\delta_4\delta_6 |-z\rangle_1|-z\rangle_2|-z\rangle_3|-z\rangle_4|-z\rangle_5|-z\rangle_6
\end{aligned} \quad (7\text{-}1)$$

All of the above states are collected in the LGN.

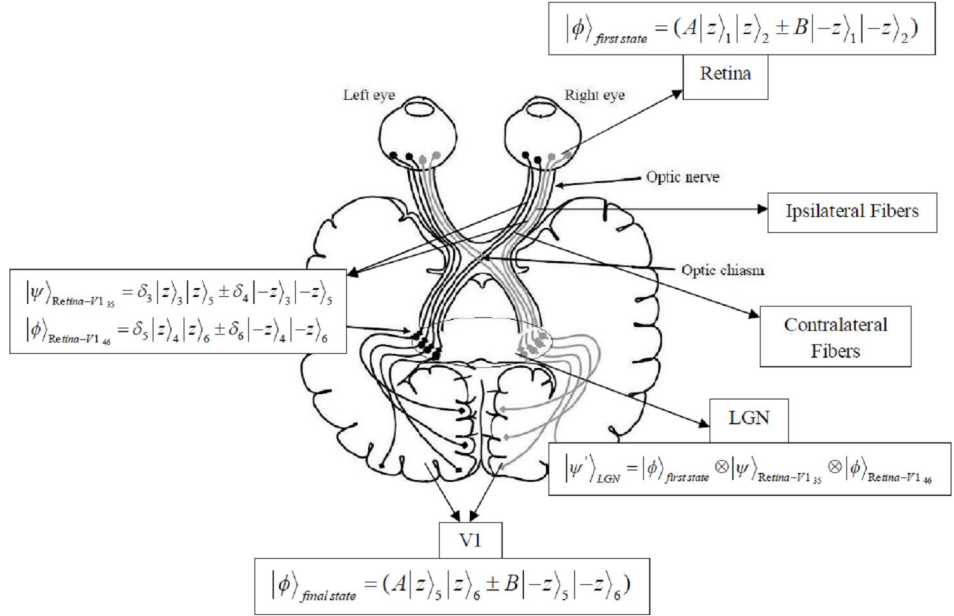

Figure 3- Representation of the teleportation model through visual pathways in the human brain. Superposed photons are entered into the eye and then absorbed (collapsed) in the retina and produced action potentials (classical signals). The entanglement and coherence between the retina and V1 are created by MTs in the visual pathways. Contralateral fibers are longer than ipsilateral fibers which causes phase shift between the two above mentioned fibers. Phase shift can reproduce the first quantum state in V1.

But here the role of action potentials is very important. They determine which fibers are fired. If fibers i and ii carry action potentials, then it shows that information passes through ipsilateral fibers. Thus to select information from the LGN to send it via fibers v and vi to V1 there is no need for phase difference (or to apply the phase shift operator on the states) and thus the state of (3-2) can be transferred like its first state through fibers v and vi. Hence,

$$i \text{ and } ii \text{ firing} \xrightarrow{yields} |\phi'\rangle_{final\,state} = A|z\rangle_5|z\rangle_6 \pm B|-z\rangle_5|-z\rangle_6 \quad (7\text{-}2)$$

In another state, if fibers i and iv are fired, it means that one fiber is selected from ipsilateral fibers and the other is from contralateral fibers, thus they have a phase difference with respect to each other. Hence,

$$i \text{ and } iv \text{ firing} \xrightarrow{yields} |\phi'\rangle_{final\,state} = A|z\rangle_5|-z\rangle_6 \pm B|-z\rangle_5|z\rangle_6 \quad (7\text{-}3)$$

To restore initial information, the operator

$$R(\varphi) = \exp(-i\pi a_6^\dagger a_6)$$

should operate on the state in LGN in which fibers i and iv have conveyed action potentials. This operator changes the ket $|z\rangle_6$ to $|-z\rangle_6$ and vice versa. It means that fiber iv has a π radian phase difference with respect to fiber i, and this phase difference can restore the exact state of the photon. If fibers ii and iii are fired, this yields:

$$ii\ and\ iii\ firing \xrightarrow{yields} |\phi'\rangle_{final\ state} = A|-z\rangle_5|z\rangle_6 \pm B|z\rangle_5|-z\rangle_6 \qquad (7\text{-}4)$$

In this case the operator

$$R(\varphi) = \exp(-i\pi a_5^\dagger a_5)$$

should be involved. For the case of iii and iv firing, this yields,

$$iii\ and\ iv\ firing \xrightarrow{yields} |\phi'\rangle_{final\ state} = A|-z\rangle_5|-z\rangle_6 \pm B|z\rangle_5|z\rangle_6 \qquad (7\text{-}5)$$

in which case the operator

$$R(\varphi) = \exp(-i\pi(a_5^\dagger a_5 + a_6^\dagger a_6))$$

should be involved. In this case we see that the main path is that of ipsilateral fibers which are directly connected to each eye and fibers iii and iv both have a $\pi$ radian phase difference with respect to it. We also know that there are two LGNs and the left and right V1 (see Fig. 3). Now, another question emerges: "how do these two left and right parts in V1 can instantaneously receive information?" To answer this question, we have to find the reason why large bundles of MTs in different neurons of cortical areas can be connected with each other. We propose that the synaptic β-neurexin/neuroligin-1 adhesive protein complex [56] besides engaging in direct interaction with MTs via gap junctions [80] can play the role of a device mediating coherence between the cytoskeletons of the cortical neurons. Neuroligin-1 is a member of brain specific family of cell adhesion proteins [56]. Indeed information in the neurons is transmitted at synapses to other neurons. It is discovered that neuroligin-1 is specifically localized to synaptic junctions [81]. The extracellular part of neuroligin-1 binds to another group of cell adhesion molecules, the β-neurexins. This β-neurexin-neuroligin junction is formed at the initial site of contact between a presynaptic axon terminal and its target cell. The intrasynaptic β-neurexin-neuroligin-1 adhesion can be seen as not only organizing the pre- and post- synaptic architecture but causing cytoskeletons of two or more neurons which act as a 'unit' system [82]. Thus it is possible the macroscopic coherent quantum state extends through large brain cortical areas.

We see that crossing of neurons in the visual pathways plays an important role in restoring information in the brain. In Fig. 3 it is seen that the contralateral fibers of the left eye cross at the optic chiasm and are connected to the right side of the brain and the contralateral fibers of the right eye cross to the left side of the brain. This crossing causes a phase shift between direct neurons and crossed neurons. It is conceivable that rotations or crossings of neurons throughout the body are there for this very reason.

## 8) Discussion and Conclusions

In general, we can briefly summarize our approach by listing the following properties:

1- We have combined the main assumptions of the Orch-OR model with Tegmark's approach and the Thaheld conclusion in a compact physical model which we call "The Teleportation Model".

2- Our model investigates visual pathways from atomic to macroscopic scales. This approach includes classical descriptions and offers new answers to open questions.
3- The proposed model explains why the shape of action potentials stays the same. Classical models state that "sensations" are action potentials that reach the brain via sensory neurons, and "perception" is the awareness and interpretation of the sensation. It is reasonable to assume that the constant shape of action potentials cannot result in different profiles of information. Thus the shape of information should be due to neurons. In this approach MTs are the representatives of information carriers. In our approach action potentials just determine which neurons fire and which do not.
4- The teleportation hypothesis explains why neurons cross at some point. This crossing causes a phase shift relative to a special pathway. In teleportation of entangled coherent states the phase shift operators can rebuild initial information.
5- Our model can describe how different information can be simultaneously perceived as a binding nature of conscious experience. This can be done via quantum parallel processing.
6- It explains how the brain of the observer can receive quantum information from the environment.

We can see that there still exists the possibility that the mind can play the main role in the measurement problem, and this is in accord with what von Neumann, London, Bauer, and Wigner (initially) asserted.

In conclusions, in this paper we have theoretically demonstrated the plausibility of a quantum teleportation mechanism between the eye and the brain which can describe different aspects of the visual processing through visual pathways. Our model is brought to bear on both quantum and classical aspects of neuroscience. It is interesting to note in closing that in a recent paper Koch and Hepp [83] grappled with this problem in an essay and concluded that when an observer looks at a quantum system like Schroedinger's cat, the quantum state of the system interacts with retina (i.e. a classical system) and collapses into just either a dead state or an alive state. Thus, these authors believe that quantum mechanics is not applicable to the functioning of the brain. We disagree with this conclusion and posit that even with a collapse of the quantum state in the retina, the brain can collapse quantum states as well. Our paper was aimed at demonstrating in terms of rigorous physical arguments how this can happen.

*Acknowledgements*
 V. Salari and M. Rahnama thank Dr. Vahid Sheybani, the head of Neuroscience Research Center, for helpful and informative discussions.  J.A. Tuszynski's research has been supported by NSERC, the Allard Foundation and the Alberta Cancer Foundation.